\def \AAP #1 #2 {{\em Astron. Astrophys.\/} {\bf #1}, #2}
\def \AAL #1 #2 {{\em Astron. Astrophys. Lett.\/} {\bf #1}, L#2}
\def \AAR #1 #2 {{\em Astron. Astrophys. Rev.\/} {\bf #1}, #2}
\def \AAS #1 #2 {{\em Astron. Astrophys. Suppl. Ser.\/} {\bf #1}, #2}
\def \AJ #1 #2 {{\em Astron. J.\/} {\bf #1}, #2}
\def \ANNREV #1 #2 {{\em Ann. Rev. Astron. Astrophys.\/} {\bf #1}, #2}
\def \APJ #1 #2 {{\em Astrophys. J.\/} {\bf #1}, #2}
\def \APJL #1 #2 {{\em Astrophys. J. Lett.\/} {\bf #1}, L#2}
\def \APJS #1 #2 {{\em Astrophys. J. Suppl.\/} {\bf #1}, #2}
\def \APSS #1 #2 {{\em Astrophys. Space Sci.\/} {\bf #1}, #2}
\def \ASR #1 #2 {{\em Adv. Space Res.\/} {\bf #1}, #2}
\def \BAIC #1 #2 {{\em Bull. Astron. Inst. Czechosl.\/} {\bf #1}, #2}
\def \JSQRT #1 #2 {{\em J. Quant. Spectrosc. Radiat. Transfer\/} {\bf #1}, #2}
\def \MN #1 #2 {{\em Mon. Not. R. Astr. Soc.\/} {\bf #1}, #2}
\def \MEM #1 #2 {{\em Mem. R. Astr. Soc.\/} {\bf #1}, #2}
\def \PLR #1 #2 {{\em Phys. Lett. Rev.\/} {\bf #1}, #2}
\def \PASJ #1 #2 {{\em Publ. Astron. Soc. Japan\/} {\bf #1}, #2}
\def \PASP #1 #2 {{\em Publ. Astr. Soc. Pacific\/} {\bf #1}, #2}
\def \NAT #1 #2 {{\em Nature\/} {\bf #1}, #2}
\def \SAIT #1 #2 {{\em Mem.\ Soc.\ Astron.\ It.\/} {\bf #1}, #2}
\def \MESS #1 #2 {{\em The Messenger\/} {\bf #1}, #2}
\def \ASTRNACH #1 #2 {{\em Astron. Nach.\/} {\bf #1}, #2}
\def\be{\begin{equation}}
\def\ee{\end{equation}}
\def\lsim{\lower 2pt \hbox{$\, \buildrel {\scriptstyle <}\over
         {\scriptstyle \sim}\,$}}
\newcommand\gsim{\buildrel > \over \sim}

\documentstyle{BSAXPwork}
\input epsf.sty
\begin{opening}
\title{High Energy Emission from the Polar Cap: The Slot Gap Revisited}
\author{A. K. Harding$^1$ \& A. G. Muslimov$^2$}
\institute{$^1$NASA Goddard Space Flight Center, Greenbelt, MD 20771, USA\\
$^2$ManTech International Corp., Lexington Park, MD 20653, USA}
\date{} 
\end{opening}

\begin{document}

\oddpagefooter{}{}{} 
\evenpagefooter{}{}{} 
\medskip  

\begin{abstract}
The characteristics of the high-energy emission from polar cap accelerators will be discussed.  Particles accelerated in the ``slot gap" near the polar cap rim will reach altitudes of several stellar radii before initiating pair cascades, producing a wide hollow cone of emission in young pulsars and some millisecond pulsars.  Model X-ray and gamma-ray spectra and pulse profiles, based on Monte-Carlo simulations of polar cap pair cascades, will be presented. 
\end{abstract}

\medskip

\section{Introduction}

The number of rotation-powered pulsars with detected emission at X-ray and $\gamma$-ray energies has been steadily growing (Thompson 2001, Becker \& Aschenbach 2002).  Although 
the theory of pulsar acceleration and high-energy emission has been studied for over twenty-five years, the origin of the pulsed non-thermal emission is a question that remains unsettled.
The observations to date have not been able to clearly distinguish between an emission site
at the magnetic poles (Daugherty \& Harding 1996) and emission in the outer magnetosphere (Cheng, Ho \& Ruderman 1986, Hirotani \& Shibata 2001).   In the case of polar cap (PC) models, while the energetics and pair-cascade spectrum have had success in reproducing the observations, the predicted beam size of 
radiation emitted near the neutron star (NS) surface is too 
small to produce the wide pulse profiles that are observed (Thompson et al. 2001).   However, Arons (1983) first noted the possibility of a high-altitude acceleration region or ``slot gap" (near the PC rim, based on the finding of Arons \& Scharlemann (1979, hereafter 
AS79) that the pair formation front, above which the accelerating field is screened, occurs at increasingly higher altitude as the magnetic colatitude approaches the last open field line where the electric field vanishes.  We have re-examined the slot-gap model with the inclusion of two new features: 1) the acceleration due to inertial-frame dragging (Muslimov \& Tsygan 1992) and 2) the additional decrease in the electric field near the boundary at the edge of the polar cap due to the narrowness of the slot gap.  These two effects combine to enable acceleration to altitudes of several stellar radii in the slot gap at all azimuthal angles around the polar cap.  These features result in the production of a larger high-energy emission beam at smaller magnetic inclination angles, both necessary for widely spaced double-peaked pulse profiles.

\section{Formation of the Slot Gap}

In the space-charge limited flow acceleration model (Arons \& Scharlemann 1979), electrons are freely emitted from the neutron star surface near the magnetic poles and accelerated in steady-state.  When the electrons achieve a sufficient Lorentz factor they radiate curvature and inverse Compton photons that can produce electron-positron pairs in the strong magnetic field.  Some of the positrons turn around and accelerate downward toward the neutron star, while the electrons accelerate upward.  This pair polarization screens the electric field above a pair formation front (PFF). These models assume a boundary condition that the accelerating electric field and potential are zero at the last open field line.  Since near the boundary, a larger distance is required for the electrons to accelerate to the Lorentz factor needed to radiate photons energetic enough to produce pairs, the PFF occurs at higher and higher altitudes as the boundary is approached.  The PFF thus curves upward, approaching infinity and becoming asymptotically tangent to the  last open field line.  If the electric field is effectively screened above the PFF, then a narrow slot surrounded by two conducting walls is formed (see Figure 1).   Within this slot gap, the electric field is further screened by the presence of the second conducting boundary and acceleration occurs at a reduced rate.  Pair production and pair cascades therefore do not take place near the neutron star surface in the slot gap, as do the pair cascades along field lines closer to the magneic pole (core), but occur at much higher altitudes.

\begin{figure}  

\epsfysize=10cm 

\hspace{2.0cm} \vspace{0.0cm}
\epsfbox{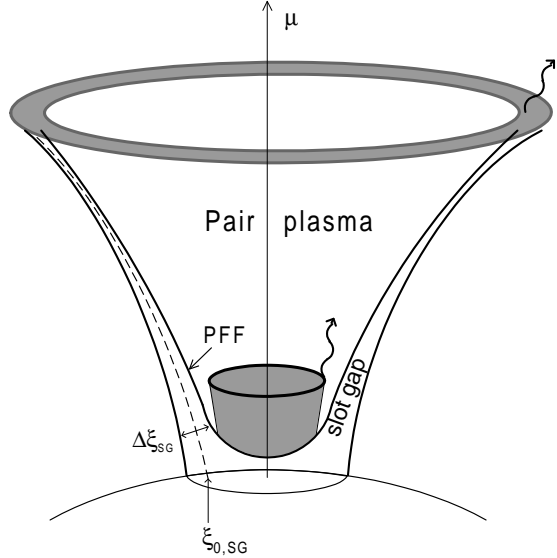}

\caption[h]{
Schematic illustration of polar cap geometry, showing the outer boundary of the open field line 
region (where $E_\parallel = 0$) and the curved shape of the pair formation front (PFF) which 
asymptotically approaches the boundary at high altitude.  The slot gap  exists between the pair plasma which results from the pair cascades above the PFF and the outer boundary.  A narrow
beam of high-energy emission originates from the low-altitude cascade on field lines interior 
to the slot gap.  A broader, hollow-cone beam originates from the high-altitude cascade above
the interior edge of the slot gap. $\Delta \xi_{_{\rm SG}}$ is the slot gap thickness (see text) and 
$\theta_{0, SG}$ is the colatitude at the center of the slot gap.
}
\end{figure}  

There are several important differences between our revised slot-gap model and the original slot-gap model of Arons \& Sharlemann (1979).  The inclusion of general relativistic frame dragging enables particle acceleration on both ``favorably" and ``unfavorable" curved field lines and also at all inclination angles.  We also consider the radiation from pair cascades occurring along the interior edge of the slot gap.  The cascade radiation emission beam from the slot gap is thus a full hollow cone centered on the magnetic axis.  A narrower emission beam from field lines interior to the slot gap will form a core component of pairs and high-energy emission.

\section{Slot Gap Energetics}

\begin{figure}   

\epsfysize=9cm 

\hspace{5.0cm} \vspace{0.0cm}
\epsfbox{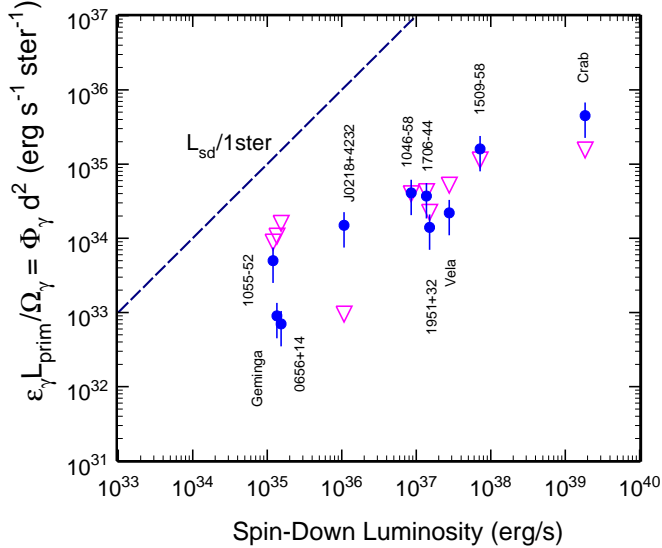}

\caption[h]{
Observed flux above 1 keV, $\Phi_{\gamma}$, times distance squared (from Thompson (2001))
(solid circles) and theoretical values of specific high-energy luminosity from the slot gap, $\varepsilon _{\gamma }~L_{\rm prim}/ \Omega _{\gamma}$ from eq. (\ref{L/Omega1}) 
(upside-down triangles) vs. spin-down luminosity for known $\gamma$-ray pulsars.  An
efficiency of $\varepsilon_{\gamma} = 0.3$ was assumed. Also $\lambda = 0.1$, $\eta_{\gamma} = 3$ 
and the stellar parameters $R_6=1.6$ and $I_{45}=4$ were used. 
}
\end{figure}

The electrodynamics of the slot gap is primarily dependent on a single parameter, the slot gap width, $\Delta \xi_{_{\rm SG}}$.   The ratio of the electric field in the slot gap to the electric field in the core region of the PC is $E_{_{\rm SG}} / E_{\rm core} \propto \Delta \xi_{_{\rm SG}}^2/4$ and the luminosity of particles accelerated in the slot gap is $L_{\rm prim} \propto \Delta \xi_{_{\rm SG}}^3 L_{\rm SD}$ where  $L_{\rm SD}$ is the spin-down luminosity of the pulsar (for full details see Muslimov \& Harding 2003).  
One can estimate the width of the slot gap as defined by the magnetic colatitude where the variation of the curvature radiation PFF becomes comparable to a fraction $\lambda$ of the stellar radius, or
\be  \label{SGwidth}
\Delta \xi _{_{\sc SG}} \approx
\left\{
\begin{array}{lr} 
 0.2~P_{0.1}(\lambda B_{12})^{-4/7}I_{45}^{-3/7}.
\label{deltaxi1} &  B \lsim 0.1 B_{\rm cr} \\
0.1~P_{0.1}(\lambda B_{12}^{3/4})^{-4/7}
I_{45}^{-3/7} & B \gsim 0.1 B_{\rm cr}
\end{array} 
\right.
\ee
where $P_{0.1} = P/0.1$ s, $B_{12} = B_0/10^{12}$ G, and $I_{45} = I /10^{45}\,\rm g\,cm^2$, are the NS rotation period, surface magnetic field and moment of inertia.
The emission solid angle of radiation from the slot gap can be estimated by integrating over the thin annulus defined by the slot gap width (Eqn [\ref{SGwidth}]).  
\be \label{Omega}
\Omega _{\gamma } \approx {9\over 2}\pi \theta _0^2 \eta \Delta \xi _{_{\sc SG}}~~~~{\rm ster},
~~~~~\eta \approx \eta _{_{\gamma }} .
\ee
where $\eta \equiv r/R$ is the dimensionless radius of emission and $\theta_0 = \sin^{-1}(2\pi R/Pc)^{1/2}$ is the PC half-angle.  Based on the luminosity of the primary electrons and the above solid angle estimate, we may derive the quantity, 
\be \label{L/Omega1}
L_{\gamma }(\Omega _{\gamma }) = {{\varepsilon _{\gamma }~
L_{\rm prim}}\over {\Omega _{\gamma}}} = 
3\times 10^{34} ~\varepsilon _{\gamma }~ 
L_{\rm sd, 35}^{3/7} P_{0.1}^{5/7} R_6^{17/7} 
\Lambda (\eta _{_{\gamma }})
~~~~~{\rm erg \cdot s^{-1} \cdot ster ^{-1}} .
\ee
where $R_6 = R/10^6$ cm is the neutron star radius, $\varepsilon_{\gamma}$ is the radiation efficiency and $L_{\rm sd,35} \equiv L_{\rm sd}/10^{35}\,\rm erg s^{-1}$ is the spin-down luminosity.
The above expressions for $L_{\gamma }(\Omega _{\gamma })$ are equivalent to the observed quantity 
$\Phi _{\gamma }~d^2$, where $\Phi _{\gamma }$ is the high-energy bolometric flux observed at the Earth, and $d$ is the distance to the pulsar. 
Figure 2 shows the pulsar empirical 
(solid circles with error bars) and theoretical (upside-down triangles) values of $\Phi _{\gamma }d^2$ as a function 
of spin-down luminosity, $L_{\rm sd}$. The theoretical values are calculated for the parameters $\varepsilon _{\gamma } = 0.3$ and $\lambda = 0.1$ (see eq. 
[\ref{L/Omega1}]). Note that parameter $\varepsilon _{\gamma }$ can 
range from 0.2 to 0.5 in cascade simulations, and $\eta _{\gamma } = 3$. 
In Figure 2 the dashed line represents the absolute upper 
limit, where the spin-down luminosity is radiated into the unit 
solid angle, i.e. where 
$\Phi _{\gamma }d^2 = L_{\rm sd} / 1~{\rm ster}$. One can see that there is good agreement for most
high-energy pulsars except several of the pulsars, Geminga and PSR B0656+14, having low $L_{\rm sd}$, and for J0218+4232, which is a millisecond pulsar.   These pulsars are near or below the death line for pairs from curvature radiation (see Harding \& Muslimov 2002, and next Section), and therefore have either very wide slot gaps or no slot gaps at all. 
All other high-energy pulsars depicted in Figure 2 
are above their curvature radiation death lines and are expected to have slot gaps.  

\section{Geometry of Emission and Pulse Profiles}

\begin{figure} 

\epsfysize=12cm 

\hspace{2.0cm} \vspace{0.0cm}
\epsfbox{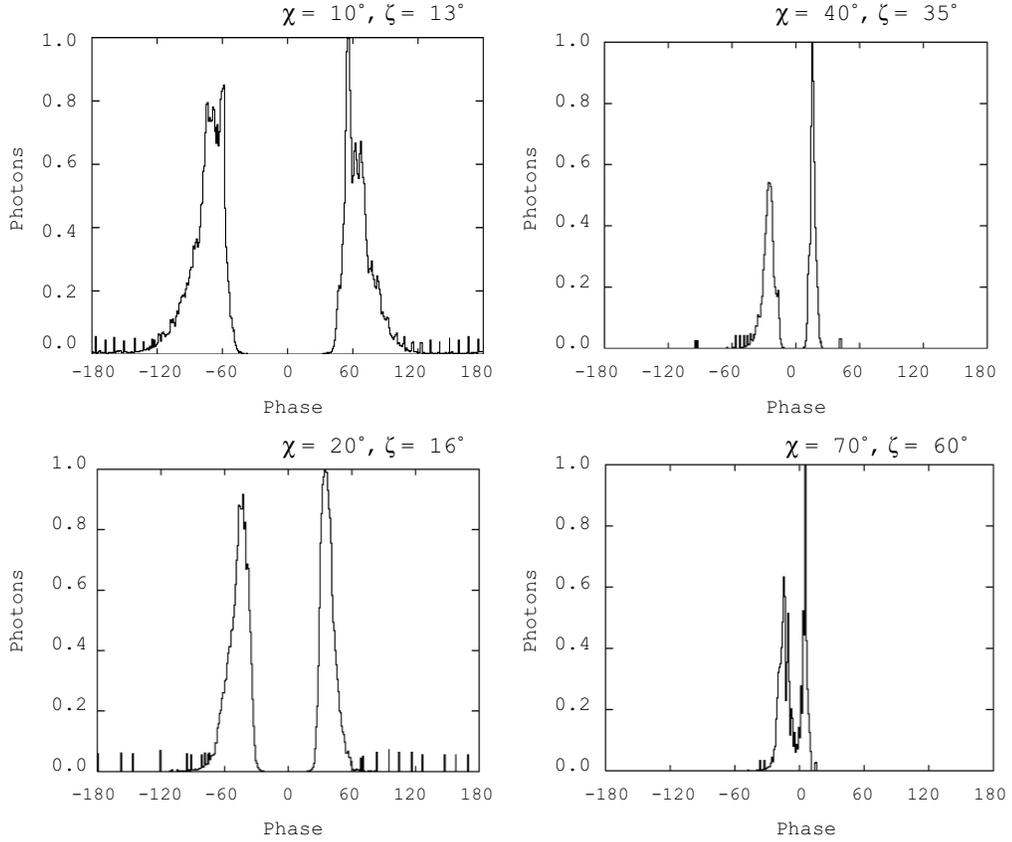}

\caption[h]{Theoretical pulse profiles of photon emission above 100 keV from the slot gap pair cascade 
at various viewing angles, $\zeta$.  The spikes on the leading and trailing edges
of the peaks in some profiles are artifacts of the resolution in magnetic phase angle around the polar cap.
}
\end{figure}

In order to explore the characteristics of high-energy emission produced by particles 
accelerated in the slot gap, we have carried out Monte Carlo simulations of the pair cascades
initiated by primary electrons accelerated by the $E_{\parallel}$ in the gap.  These pair cascade simulations are based on those of Daugherty \& Harding (1996), but using the electric field of the slot gap (see Mulsimov \& Harding 2003).  To illustrate the features of the SG acceleration and cascades, we use the period 
($P_{0.1} = 0.33$) and surface magnetic field ($B_{12} = 5$) of the Crab pulsar for the case $B < 0.1~B_{\rm cr}$. 
A stellar radius of 16 km is assumed.  Figure 3 shows examples of pulse profiles for different inclination angles $\alpha$ and seen at various viewing angles $\zeta$ to the rotation axis.   It is evident that widely spaced double-peaked profiles can result at relatively 
small inclination by slicing through the middle of the hollow cone.  Profiles with two pulses 
separated by more than 100 degrees, as are seen from the Crab,  Vela and Geminga, can be formed 
when the inclination angle is comparable to the beam opening angle (about 20 degrees in this model 
for the Crab).   Broad single pulses, as seen from PSR B1509-58, may be produced when the 
observer cuts along the edge of the cone.   As the inclination angle increases, the maximum pulse 
separation decreases and the pulses themselves also become narrower. Bridge emission as seen in 
Vela and Geminga high-energy profiles (and in the Crab profile around 1 MeV) is not produced in
this calculation since we have modeled only cascades and emission from the slot gap, which  
produces a pure hollow cone beam.  The core cascades will produce additional emission on field lines interior to the slot gap which will fill in the profile between the two pulses and produce bridge emission.

\section{Summary}

The revised picture of acceleration and radiation from pulsar slot gaps makes several observational predictions that seem to agree with some of the properties of the known high-energy pulsars.  Although the total luminosity produced by the slot gap is lower than that from the whole polar cap, the small solid angle of the slot gap allows for fluxes that are in agreement with those observed.  The opening angle of the high-energy emission beam from the slot gap is much larger than that of emission from near the neutron star surface.  This naturally allows for wider, double-peaked pulse profiles as observed in several $\gamma$-ray pulsars.  However, this type of profile still requires small magnetic inclination angles that are comparable to the emission beam opening angle.  Even for the fast pulsars, the slot gap beam opening angles are only $10^0 - 20^0$.  Applying this model to the observed $\gamma$-ray pulsars would therefore require that these, and probably most, pulsars have inclination angles smaller than $20^0$ and our viewing angle cannot be much larger.  This may be problematic in light of the Chandra X-ray image of the Crab nebula; if the plane of the torus is really orthogonal to the spin axis, then our viewing angle is $\sim 60^0$.  However, a number of other, more recently discovered, young pulsars seem to have small magnetic inclination (Harding et al. 2003) and slot-gap radiation may be visible at X-ray and $\gamma$-ray energies.


\end{document}